\documentclass[aps,twocolumn
]{revtex4}
\usepackage{amsmath}
\usepackage{amssymb}

\usepackage{graphicx}
\usepackage{bm}
\newcommand{\be}{\begin{equation}}
\newcommand{\ee}{\end{equation}}
\newcommand{\ba}{\begin{eqnarray}}
\newcommand{\ea}{\end{eqnarray}}

\newcommand{\beq}{\begin{equation}}
\newcommand{\eeq}{\end{equation}}
\newcommand{\bea}{\begin{eqnarray}}
\newcommand{\eea}{\end{eqnarray}}
\newcommand{\beas}{\begin{eqnarray*}}
\newcommand{\eeas}{\end{eqnarray*}}

\begin{document}

\bibliographystyle{apsrev} 

\title {Can the Standard Model CP violation near the $W$-bags 
 \\ explain the cosmological baryonic asymmetry?   }
\author{Yannis Burnier}
\email[Email:]{yburnier@notes.cc.sunysb.edu}
\author{Edward Shuryak}
\email[Email:]{shuryak@tonic.physics.sunysb.edu}
\affiliation{
Department of Physics, State University of New York,\\
 Stony Brook, NY 11794, USA}

    \date{\today}


    \begin{abstract} 
In the scenario of cold electroweak baryogenesis, oscillations of the Higgs field lead to metastable domains of unbroken phase  where the Higgs field nearly vanishes.
Those domains have also been identified with the $W-t-\bar{t}$ bags, a non-topological solitons made of large number ($\sim 1000$) of gauge quanta
and heavy (top and anti-top) quarks. 
As real-time numerical studies had shown, sphalerons (topological transition events violating the baryon number) occur only inside those bags.
In this work
we estimate the amount of CP violation in this scenario coming from the Standard Model, via the Cabibbo-Kobayashi-Maskawa  (CKM) quark mixing matrix, 
resulting in top-minus-antitop difference of the population in the bags. Since these tops/anti-tops are ``recycled" by sphalerons, this population difference
leads directly to the baryonic
asymmetry of the Universe.
 We look at
the effect appearing in the 4-th order in weak $W$ diagrams describing interference of different quark flavor contributions. We found that there are multiple cancellations
of diagrams and clearly sign-definite 
 effect appears  only in the 6-th order expansion over  flavor-dependent phases. We then estimate contributions
 to these diagrams in which weak interaction occurs (i) inside, (ii) near and (iii) far
from the 
the  $W-t-\bar{t}$ b-bags, optimizing the contributions in each of them. We conclude that the second (``near") scenario is the dominant one, 
producing CP violation of the order of $10^{-10}$, in our crude estimates. Together with the baryon violation rate of about $10^{-2}$, previously demonstrated for this scenario, it puts the resulting asymmetry 
close to what is needed to explain the observed baryonic asymmetry in the Universe. Our answer also has a definite sign, which apparently seems to be the correct
 one.
    \end{abstract}
    
   \maketitle 
    

    \section{Introduction}

The question how
the observed baryonic asymmetry of the Universe was produced is 
among the most difficult open questions of physics and cosmology. The observed effect is usually expressed as the ratio of the baryon
density to that of the photons $n_B/n_\gamma\sim 10^{-10}$. 
Sakharov \cite{Sakharov}
had formulated three famous necessary  
conditions:  the (i) baryon number and (ii) the CP violation, with (iii) obligatory deviations from the thermal equilibrium.  
Although all of them are present in the Standard Model (SM) and standard Big Bang cosmology,
 the baryon asymmetry which is produced by the known CKM matrix is completely insufficient to solve this puzzle.

Significant efforts have been made to solve it using hypothetical  ``beyond the  standard
model" scenarios, for instance related with possible large CP violating processes in the neutrino mass matrix or in the supersymmetric sector.
  Another alternative
  is the modification of the standard cosmology.
While the standard Big Bang scenario predicts adiabatically slow  
crossing of the electroweak phase transition, leading to extremely small deviations from equilibrium,  
the so called ``hybrid" or ``cold" scenario  
\cite{GGKS,KT,Felder:2000hj,GarciaBellido:2002aj}   leads to 
large deviations  from equilibrium. This scenario
combines the end of the inflation era with the establishment of
the  electroweak broken phase,
 avoiding some pitfalls of the standard  cosmology, such as an ``erasure" of asymmetries generated before the electroweak scale
 by large sphaleron rates in the symmetric (electroweak plasma) phase. 

Studies of this scenario in the last decades have been rather intense.
Coherent oscillations of the gauge/scalar fields   have been studied in detail in  real-time lattice simulations \cite{GarciaBellido:2003wd,Tranberg:2003gi,Tranberg:2009de, Tranberg:2010af}. 
They have revealed relatively long lived ``hot spots" with depleted vacuum expectation value (VEV) of the Higgs field.
They have further found that all sphaleron transitions take place only $inside$ these spots. The rate  
of transitions in the symmetric (no Higgs VEV) phase was found to be many orders of magnitude larger \cite{Moore:2000mx, Moore:2010jd} than via standard electroweak sphalerons, even including modifications
near the phase transition \cite{Arnold:1987mh, Arnold:1987zg, Burnier:2005hp}. Qualitative explanation of the metastability of these hot spots has been provided by finding metastable bags filled with gauge bosons and 
top quarks \cite{Crichigno:2010ky}.  The enhanced sphaleron were explained analytically using by the so called COS sphalerons \cite{Ostrovsky:2002cg}, which have
significantly larger sizes and thus smaller masses
than the standard sphalerons in the broken phase, see details in \cite{Flambaum:2010fp}. The amount of the baryon number violation in this scenario
can reach $10^{-3}$, or  even $10^{-2}$ with the top quark ``recycling" mechanism \cite{Flambaum:2010fp}.  

This paper is devoted to evaluation of the  CP-odd asymmetry resulting in the SM from the well known CKM matrix. Its manifestations
has been first discovered in Kaon decays and lately studied extensively
in decays of the $B$ mesons,  providing
all elements of the  CKM matrix, with accuracy described in the current Particle Data Tables. The first attempts to estimate  magnitude of CP violation
 in cold electroweak cosmology has been made by Smit, Tranberg and collaborators \cite{Hernandez:2008db, Tranberg:2009de}. 
Their strategy has been to derive some local effective CP-odd Lagrangian by integrating out quarks, such as that found in \cite{Hernandez:2008db},
and than include this Lagrangian in their real-time bosonic numerical simulations. The estimated magnitude of the CP-odd effects ranges from $n_B/n_\gamma \sim 10^{-6}.. 10^{-10}$ \cite{Tranberg:2010af}, which reignites hopes that this scenario can provide the observed magnitude of the baryon asymmetry in Universe.

However, 
as we will detail below, there are many unanswered questions about the accuracy of these estimates. One of them, already pointed out e.g.~in \cite{Flambaum:2010fp}, is that the particular effective Lagrangian has been derived with specific assumptions about the scale of the loop momenta and field strength, which are only valid in some restricted regions in the configurations used for averaging. The effective theory used,  valid e.g. for energy scales below $5$ GeV \cite{Hernandez:2008db} cannot be
applied in and near the bags  of electroweak  size $1/m_w$: and yet their large CP effect came precisely  from these regions. 
But even more important is the following unanswered generic question:  why should a very complicated operator (containing 4-epsilon symbol convoluted with 4 gauge field potentials and one field strength) averaged over very complicated field configurations (obtained only numerically) have $a$ nonzero average?
Since the calculation is numerical, it would be desirable to have some parametric estimate of the effect, in particular to know
what sign the effect should have and why. 

In this work we will try to elucidate these issues by evaluating CP-odd effects induced at the one quark loop level in the W-bag background.  We will find which quarks and which scales interact with the fields of the $W$-bags, and estimate the value of the CP asymmetry produced.
We will find many cancellations in the contributions of various quark flavors in the loops, with our \emph{definite sign} answer only coming in the 6th order expansion over
the propagator phases. Instead of a common scale for all propagators, as is used in the effective Lagrangians, the positions of the interaction points
are individually optimized to maximize the effect. 
In section \ref{s2} we review the properties of the cold electroweak phase transition and the formation and properties of the bags. We then explain how fermions can built CP asymmetry. In section \ref{s3} we write down the CP asymmetry and decide, which quark interference can create the CP asymmetry by working out the flavor algebra. Working out the propagators and Dirac algebra allow us finally to estimate the size of the different effects in section \ref{se}. In section \ref{s4} we discuss and summarize our results.

\section{The setting} \label{s2}

The cold electroweak phase transition is completely out of thermal equilibrium. The first SM fields created by violent fluctuations
 are those of the $W,Z$ and Higgs bosons, the latter
promptly producing  top and anti-top quarks (due to their large Yukawa coupling). Numerical simulations \cite{Tranberg:2009de, Tranberg:2010af} of the gauge and Higgs fields have shown that $W$-bags of size of order $1/m_w$ are formed, containing large number of gauge quanta.
Top quarks and antiquarks are migrating into these bags, where their large mass is nearly canceled by a large binding of  $O(100\, GeV)$, as shown 
in \cite{Flambaum:2010fp}. The bag lifetime is  of order $6/m_w$ and the system finally thermalize, to some equilibrium temperature well below the electroweak critical temperature.

Note that during this time, mostly weak bosons, Higgs and top quarks are present in the system, with light quarks and gluons are not yet produced. 
This comment is important for the following reasons. Below we will evaluate certain 4-th order electroweak diagrams, whose small CP-odd phases are
interfering. In order to preserve those phases it is important that the quarks diffuse freely and don't get scattered by the thermal bath. (Note that it was one of the criticism to the mechanism proposed
 some time ago by Farrar and Shaposhnikov  \cite{FarrarShaposhnikov}.)  
In our scenario, the mean free path is of order $\lambda\sim 1/\alpha_W^2 T$. It is actually longer than all distances considered below, so that we do not need to consider thermal rescattering. The scattering by gluons is also negligible, due to their small density at early time.

Given the particle content, it is natural to search for CP violation starting form the early created top quarks. 
In the bags, they can absorb $W^-$ and turn into the down $b,s,d$ quarks. The light quarks have too small Yukawa coupling to be bound to the bag,  and can escape the bag and move into the bulk plasma. CP violation produces a difference between the rates of top and anti-top quark escape. For 
definiteness, let us consider two interactions with the $W$ bosons (returning the quark back to ``up" flavors). 
For a top quark starting  at the position $r_1$, the escaping amplitude has  the form
\beas
A_t(r_1)&=&g^2\int \gamma^\nu W^-_\nu V^\dagger P_t S_d^L(r_1,r_2) 
\\&&\times\gamma^\mu W^+_\mu V  S_u^L(r_2,r_c)d^4r_2,\notag
\eeas
whereas the antitop has
\beas
A_{\bar t}(r_1)&=&g^2\int\gamma^\nu W^+_\nu \bar{V^\dagger}P_{\bar t} S_d^L(r_1,r_2) \\&&\times\gamma^\mu W^-_\mu \bar{V}\notag  S_u^L(r_2,r_c)d^4r_2.
\eeas
In the formulas above, $g$ is the weak coupling, $V$ is the CKM matrix, $S$ the quark propagators, their index $u,d$ etc denotes the up and down quark flavors and $P_t$ denote the 
flavor projection matrix on the top quark. 

The probability of a top quark escaping from $r_1$ is then given by the integral over all positions and sum over all intermediate and
final states $f$ of the squared amplitude
\bea
Prob_t(r_1)&=&\int_{f} A_t^\dagger A_t = Tr \int d^4r_c \sum_{u} A_t^\dagger A_t \notag
\\&=&g^4\int d^4r_c d^4r_2 d^4r_3 Tr\Big[P_t\gamma^\nu W^-_\nu V^\dagger \notag
\\&& S_d^L(r_1,r_2) \gamma^\mu W^+_\mu V  S_u^L(r_2,r_c) \\&& S_u^{L\dagger}(r_c,r_3)
\left. \gamma^\alpha W^-_\alpha V^\dagger S_d^{L\dagger}(r_3,r_1)\gamma^\beta W^+_\beta V\right].\notag
\eea
Note that the interference terms between different paths are of the fourth order in the weak interactions and thus contain 4 CKM matrices. The total number difference of top quarks escaping is then
\beq
N_{t-\bar t}=\int d^4x_1 n_t(r_1)(Prob_t-Prob_{\bar t}),
\eeq
where $n_t(r_1)$ the number density of top quark in the bag (that we consider equal to the density of anti-top quarks in the first approximation)

\begin{figure}
\includegraphics[width=6.cm]{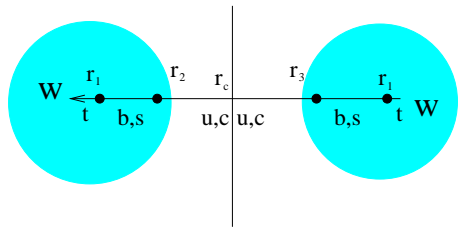}
\caption{(Color online) Schematic shape of the fourth order process involving only quarks of the 2nd and 3rd generations. The shaded objects on the left  and right represent the Higgs bag with strong gauge fields (indicated by W in the figure) inside. The vertical line is the unitarity cut. The four black dots indicate the four points $r_i,i=1..4$ where the $W$ quanta are interacting with the quark,
changing it from up to down component.} 
\label{fig_2bags}
\end{figure}

The setting in coordinate space is schematically shown in Fig.\ref{fig_2bags}. 
Four positions of the points at which the interactions take place, as well as particular quark flavor in the intermediate line, are summed over. Writing the amplitude squared
of the process, one includes the unitarity cut (the vertical line in Fig.\ref{fig_2bags}) to the right of which one, as usual, finds the conjugated image
of the process in opposite direction. Thus the interference terms have four $W$ interactions, with four CKM matrices, which is the minimal number needed for the
CP-violating effects to manifest themselves. The general expression for the amplitude will be discussed in the next section. 

In between these four points the flavor of the quark remains unchanged. Quark wave functions (we keep in mind $l=0$ or $s$-wave ones only, thus
points are only indicated by their radial distance from the bag)  are different for each flavor, because each has a different Higgs-induced potential.
Semiclassically the phase is approximated by
  \be 
  S_{12}=exp[i \int_{r_1}^{r_2}p(x)dx]\approx exp[i \int_{r_1}^{r_2}(E-{\frac{m_i^2(x) }{ 2E}}) dx],\notag
  \ee
where $E$ is the quark energy, and the approximation implies that all lower quark flavors are light with respect to $E$. If the flavor-dependent phase
  (stemming from the second term in the bracket) is small 
  \be  \delta^i_{12} =\frac{m_i^2 }{ 2E} r_{12} < 1, \label{3}\ee
we can further expand the exponent to get a series in the phases $\delta_i$.
As we will see shortly, only this  (small but flavor-dependent) part of this phase is contributing, because when two flavors produce the same answer (one gets the unit 
flavor matrix) two subsequent CKM
matrices cancel out and effectively wipe out the CP-odd part of the amplitude. 

\begin{figure}
\includegraphics[width=6.cm]{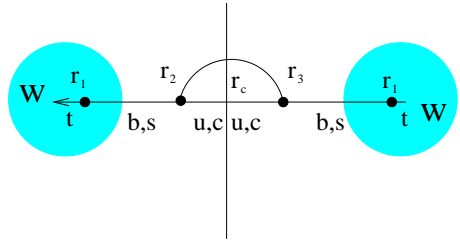}
\caption{Process where two of the $W$ boson interactions are form the bag and two $W$ are thermal but correlated.}
\label{fig_2bags2}
\end{figure}

Note that we have two possibilities, shown in Fig.\ref{fig_2bags} and \ref{fig_2bags2}. In the first case (Fig.\ref{fig_2bags}), the four interactions with the $W$ can occur 
in two pairs, in each one $W$ is $inside$ the bag and the second $near$ it. Thus in each pair one point is in the region of strong field and one in the region of the weak fields: yet they are still correlated in their spatial and SU(2) directions.
In the second case two interactions are in the bag and two occur far from it, due to presence of
 the thermal fields  outside the bags (Fig.\ref{fig_2bags2}). We will evaluate below these two possibilities subsequently.
The calculation of the escape probability $P_t$ is a formidable task in general, we will only estimate it here. We will consider that the bag and the top quark density inside is spherically symmetric and that the quark will escape radially only.

\section{CP asymmetry of the probability for the quark travel to/from the bag}\label{s3}
\subsection{The flavor algebra of the asymmetry}
  
  Let us first follow the flavor part of the amplitude that distinguishes between quarks and anti-quarks. 
  The fourth order process we outlined in the previous section corresponds to the trace of the following matrix product
  \be 
  M_t=Tr(P_t* V*S_{12}^{down}*V^\dagger*S^{up}_{23}*V*S^{down}_{31}*V^\dagger),
  \ee
where $S$ are quark propagators, the lower indices specify their initial and final points as shown in Fig.\ref{fig_2bags}, the upper subscript remind us
that those are for up or down quark components. $P_t=\mathrm{diag}(0,0,1)$ is the projector requiring that we start (and end the loop) in the bag, with a top quark.
  We also define the amplitude for the antitop quarks
  \be 
  M_{\bar{t}}=Tr(P_t* V^{*}*S_{12}^{down}*V^{T}*S^{up}_{23}*V^{*}*S^{down}_{31}*V^{T})
  \ee
which we  subtract from $M_t$, as 
the effect we evaluate is the difference in the top-antitop population inside the bag.   
The difference gets CP-odd as seen from its dependence on the CP-odd phase $\delta$
  \ba
  M_t- M_{\bar{t}}= 2 i J (S^u_{23}-S^c_{23}) \nonumber \\ 
 (-S^s_{31} *S^b_{12}-S^d_{31} *S^s_{12}+S^d_{31} *S^b_{12}   \nonumber \\ 
 +S^d_{12} *S^s_{31}-S^d_{12} *S^b_{31}+S^s_{12} *S^b_{31})
 \label{eqn_Mt-Mantit}
   \ea
  where $J$ is the so called Jarlskog factor, containing all CKM angles in the following combination
  \ba 
J=  \sin(\delta) \sin(\theta_{12})\sin(\theta_{13})\sin(\theta_{23})\cos(\theta_{23}) \nonumber \\ 
\times \cos(\theta_{12})\cos^2(\theta_{13})\cos(\theta_{23}) \approx 3*10^{-5}
  \ea
  (note that  one $cos$ is squared: this is not a misprint). 

The remaining combination of propagators, organized in two brackets, needs to be studied further. Note first that the propagators between points 2 and 3 (through the
unitarity cut) factor out and that one may ignore the top quarks there. Note further, that if the $u,c$ quarks would have the same mass, 
the first bracket would vanish: this is in agreement with general arguments that any degenerate quarks should always nullify the CP-odd effects,
as the CP odd phase can be rotated away already in the CKM matrix itself.

The last bracket in (\ref{eqn_Mt-Mantit}) contains interferences of different down quark species: note that there are 6 terms, 3 with plus and 3 with minus.
Each propagator has only small corrections (\ref{3}) coming from the quark masses. Large terms which are flavor-independent always cancel out, in both brackets of expression (\ref{eqn_Mt-Mantit}).   Let us look at only the terms which contain the heaviest $b$ quark in the last bracket,
  using the propagators in the form 
\be S^b_{ij}=\exp\left( \pm i \delta^b_{ij}  \right)=  \exp\left( \pm i \frac{m_b^2}{2 E} r_{ij} \right), \ee
where $\pm$ refers to different signs in the amplitude and conjugated amplitude and $r_{ij}=r_j-r_i$. Note that the sign of the phase between points $r_2$ and $r_3$ can be positive or negative as it results from a subtraction of the positive phase from $r_3$ to the cut $r_c$ with the negative phase form the cut $r_c$ to $r_2$. Terms containing odd powers  in $r_{23}$ should therefore  vanish in the integral and the lowest term we have is quadratic.
  Considering all phases  to be small due to $1/E$  and using the mass hierarchy $m_b \gg m_s \gg m_d$ 
  we pick up the leading contribution of the last bracket in (\ref{eqn_Mt-Mantit}) which has $r_{23}^2$
  and the 4th power in the last bracket, the 6th order in the phase shift in total:
  \be 
   M_t- M_{\bar{t}}\propto  J   \frac{ m_b^4 m_c^4 m_s^2 r_{23}^2 r_{12}  r_{31} (r_{12}+   r_{31})}{ 64 E^5}.
\label{F}
  \ee 
 Note that all distances in this expression are defined to be real and positive and the sign in the last bracket is plus, so unlike all the previous orders in the phase expansion,
 at this order we have sign definite answer with no more cancellations possible. This point is the central one in this work.
 
We further see that this expression grows for large $r$'s, which are to be integrated over. Of course as we expanded the exponent in the phases, they have to be such that these phases are smaller than 1. This means the distances are limited by $r<E/m_q^2$ and as we have a closed loop they are all smaller than $E/m_c^2$.
At larger distances powers of the phases $\delta_i$ becomes oscillating $\sim \sin\delta_i$ and may average out to zero: we would not include
these regions in our estimates below.

\subsection{Dirac algebra}
Considering the magnetic bag of \cite{Crichigno:2010ky}, with radial profile $f(r)$ and considering that the quark move radially, i.e. its propagation is described by 
\beq 
S=\frac{\gamma^\mu r_\mu}{r^4}e^{i p r}.
\eeq
Note that the phase has already been taken into account in the flavor algebra.
Only the left part of the propagators contributes in the loop since the right part does not couple to the weak fields, so that the $\gamma$-matrices can be replaced by Pauli-matrices. 
We also do not include factors from the solid angles, as they cancel between our four integrations
and propagators.

We consider first the case where all interactions are with the weak field of the bag.
With such propagators we get the Dirac algebra contribution to the amplitudes: 
\bea
\Gamma_{t}=\Gamma_{\bar{t}}&=& \frac12\notag \frac{(r_c-r_2)(r_c-r_3)(r_2-r_1)(r_3-r_1)}{(r_c-r_2)^4(r_c-r_3)^4(r_2-r_1)^4(r_3-r_1)^4}\\
&&\times f(r_1)^2f(r_2)f(r_3).
\label{D1}
\eea
In the case of two interactions with thermal fields (see Fig. \ref{fig_2bags2}), we have
\bea
\Gamma_{t}=\Gamma_{\bar{t}}&=&\frac{(r_c-r_2)(r_c-r_3)(r_2-r_1)(r_3-r_1)}{(r_c-r_2)^4(r_c-r_3)^4(r_2-r_1)^4(r_3-r_1)^4}\notag
\\&&\times f(r_1)^2 W^-_i(r_3) W^+_i(r_2).
\label{D2}
\eea
We still have here to average over the thermal fluctuations of $W^\pm$ and we approximate 
\beq
\langle W^-_i(r_3) W^+_i(r_2)\rangle \sim T \frac{e^{-m_w |r3-r2|}}{|r_3-r_2|}.\label{corrW}
\eeq
\section{Estimate of the effect} \label{se}

\subsection{Naive estimates}

Let us start with a ``naive" estimate, which assumes that $E$ in the formula is given by the
top quark mass \be E\sim m_t=173\, GeV \ee
(as all the processes of quark propagation start from tops in the bags).
As for the field strength, naively one may take all four interaction points inside the bags, where
the amplitude of the $W$ is the strongest. If so, all distances $r_{ij}$ are of the order of the bag size $R_{bag}\sim 1/m_w$.

However, if this is the case, all the phases are so small that the resulting CP asymmetry is about 10  orders 
of magnitude smaller than needed. (In fact the reader familiar with the history of the CP literature
will immediately recognize the old Jarlskog argument \cite{Jarlskog}, stated that if the scale of the process is higher than all masses,
one must always find the  product of the differences of all masses in the numerator. This is exactly what is happening in the current estimate.)

\subsection{Small quark energy and near-bag points}

However, top quarks are bound in the bag, so one may consider quark propagating  at the energy much smaller than the top mass 
$E \ll m_t$. As it has been argued in \cite{Flambaum:2010fp}, the magnitude of the weak interaction of quarks with the electroweak plasma outside the bag,
known as the screening mass, is of the  scale $\sim g_wT$, which is few GeV and  also comparable to $m_b$.
This effect is nothing but the forward scattering of a quark on electroweak plasma. 
This is the natural scale to take: thus we will from now on consider $E\sim m_b$. 

Another improvement one may try is to consider  locations of some points $outside$ the bag, selecting $r_{ij}$  as large as possible.
The escape probability is the product of the result for the flavor algebra (\ref{F}) and the Dirac algebra (\ref{D1}, \ref{D2}). In the first case, putting together the formula (\ref{F}, \ref{D1}), shifting the integration by $r_1$, ($r_i\to r_i+r_1$) we can integrate over $r_c$. The result can be approximated by noting that from the factor $(r_3-r_2)^2$ in (\ref{F}) the integral is dominated by configurations with $r_2\ll r_3$ or symmetrically, so that the result can be simplified to
\beas
Prob_{t-\bar t}(r_1)&\sim&Jg^4\frac{m_b^4m_c^4 m_s^2}{64 E^5}\int dr_2 dr_3 2 r_2 r_3^2\\ &&\times f(r_1)^2 f(r_2) f(r_3).
\eeas
Considering a radial bag of $N_w$ weak bosons, we approximate the bag shape by an exponential profile. Following Ref. \cite{Flambaum:2010fp} we get
\beq
W(r)=\sqrt{\frac{N_w m_w^3}{\pi E_w}}e^{-m_w r},
\eeq
where $E_w$ is the energy of a $W$ Boson in the bag ($E_w\sim m_w/2$).
The exponential fall-off limits the distance to which the quarks can travel and we get that the probability of a top quark escaping is
\beq
\delta_{CP}=\tau J  \frac{g^4 N_w^2 m_c}{16 E}\frac{m_b^4}{E^4}\frac{m_c^3}{m_w^3}\frac{m_s^2}{E_w^2}\sim 10^{-11}\left(\frac{N_w}{1000}\right)^2.
\label{best}
\eeq
In the latter formula we made use of the lifetime of the bag denoted $\tau/m_w$, with $\tau\sim 6$ to bound the time integral over $x_1$. We also considered that the energy of the initial top quark bound to the bag was of order $E\sim m_b$.
The main reason for the result to be small is the small radius of the bag $\sim 1/m_w$. Even in the time direction, the lifetime of the bag is small. 

\subsection{Two weak interactions far from the bag }

One can try to increase $r_{ij}$ even further, since the tail of the W fields of the bag would eventually
dive into the thermal sea of the electroweak plasma background, nonzero at any distance.
In this case the field 
strength is defined by the ``outside temperature" $T$ far from all bags. This temperature is expected to be 
below electroweak critical temperature $T<T_c$, in numerical simulations it was $T\sim 50\, GeV\sim (1/2)T_c$. 

However, the fields in the plasma are chaotic, and 
the correlator (\ref{corrW}) of two gauge fields  decreases fast with distance. Thus two points $r_2$ and $r_3$ in this case have to be sufficiently close to each other. From (\ref{F}, \ref{D2}), shifting the integration by $r_1$, ($r_i\to r_i+r_1$) and expanding around $r_{23}=0$, we get:
\bea
Prob_{t-\bar t}(r_1)&\sim&J g^4 T\frac{m_b^4m_c^4 m_s^2}{64 E^5}\int dr_2 dr_3 \frac{1}{20}(r_2+r_3)\notag\\ &&\times(r_2-r_3)^2 e^{-m_w|r3-r2|} f(r_1)^2.
\eea
Changing again the integration variable $r_3\to r_3+r_2$ we can perform the $r_3$ integral and finally the $r_2$ integral is to be bounded by the life time of the bag that we denote $\tau/m_w$ (and the distance to which quarks can travel bound to $d< c \tau/m_w$), leading to
\beq
Prob_{t-\bar t}(r_1)\sim J g^4 T\tau^2\frac{m_b^4m_c^4 m_s^2}{64 E^5 m_w^2}f(r_1)^2.
\eeq
The probability of a top-minus-antitop quark escaping from the bag is then
\beq
J g^4 \frac{\tau^3 T N_w}{ 640 E_w}\frac{m_b^4m_c^4m_s^2}{E^5 m_w^5}\sim 10^{-15} \left(\frac{N_w}{1000}\right),\notag
\eeq
which is smaller  that in the previous case. Again, the quarks, even if not bound to stay into the bag cannot propagate very far, due to the short lifetime of the bags.

\section{ Translating the results into baryon asymmetry  } \label{s4}
Let us start the summary by reminding the reader why
 evaluation of the CP-violating effects in the cosmological setting is technically so difficult.
One general reason for it is that one cannot use standard ``effective Lagrangian" method, in which the loop
momentum scale is large compared to all masses: as shown by Jarlskog long ago, this produces simple answer which however imply negligible CP violation $
\sim 10^{-20}$ . If one uses larger scale for the loop momenta, like $m_c$ or $m_b$
 the result increases, as found e.g. in Refs \cite{Tranberg:2009de, Tranberg:2010af}.  However 
such resulting Lagrangian is a very nonlocal object, and  it is not clear how one can get any reliable estimates based on them
for complicated fields obtained in numerical simulations. In particularly, as we already commented before, it is completely inadequate
for the ``bags" themselves.
 
The main lesson we got from this study
is that the scales of both the quark energy $E$ and their traveling distances $r_{ij}$ in the loop amplitudes should be tuned individually,
 to maximize the effect. The main limitation come from the conditions of quark rescattering in  the plasma (the screening masses)
 and the conditions that all phases $\delta_i$ should not be large,
 as well as the limitations coming from the $W$ field strength and correlation length.
 Another lesson is that in order to prevent cancellations between different flavors, one has to expand all the
 results till sign definite answer is guaranteed.   
    
 Is the largest CP effect we found, given in (\ref{best}),  in the right ball park for the cosmological baryogenesis? 
To answer this question we have to reprocess the CP asymmetry obtained into the baryon number.
It was shown in a toy model \cite{Burnier:2006za} that the presence of heavy quarks accelerate the sphaleron rate such as to destroy them. We will not attempt to calculate the sphaleron rate and the influence of the CP asymmetry on it but rely on the results of the gauge field simulations of Ref. \cite{Tranberg:2009de}, according to which the efficiency of the CP asymmetry conversion to baryonic asymmetry 
is found there to be of order $10^{-3}$.
Presence of tops inside the bag makes their ``recycling"  possible, and it was argued in \cite{Flambaum:2010fp} the baryon number increased to about $10^{-2}$ due to using their mass for barrier penetration. What it means is that instead of standard SM sphaleron production of $0\rightarrow 12$ fermions, 3 leptons and 9 quarks,
 one may use a process which uses top quarks of all colors e.g. $3\bar{t} \rightarrow 9$, which is favored since it require less energy. 
The probability to find 3 antitops is actually proportional to $(1+\delta_{CP})^3\approx 1+3\delta_{CP}$,
 while it is $ (1-3\delta_{CP})$ for tops: it gives factor 3. Another factor
 3 appears because of the fact that each sphaleron event creates 3 units of baryon number, not one. 
 Together with baryon asymmetry (time integrated) sphaleron rates of $10^{-2}$ and $3\times3\times\delta_{CP}$ we arrive to our final estimate
 \be 
 \frac{\Delta B}{B+\bar B}\sim 10^{-12\pm 1}
 \ee   
where $B$ and $\bar B$ are the density of baryons and anti-baryons in the system, mostly the top quarks in the bags and $\Delta B=B-\bar B$ the baryon asymmetry. The one order of magnitude stands for our errors due to numerical factors ignored in the estimates. Note that the final baryon asymmetry would still receive a small suppression due to the entropy release at the late stage of the universe expansion. We conclude that
it is somewhat below the observed baryonic asymmetry of the Universe, however parameters of the cosmological model can be better tuned to get closer the right value. For instance if it is possible to obtain larger bags with $N_w\sim 10000$ or with larger radius, the scenario might rapidly become viable.

Last but not least is the issue of the $sign$ of the asymmetry. Our formula (\ref{F})  has definite (positive) sign, that is to say more top quark escape the bag (note that the time direction is important, quarks are first created in the bag, then have more probability to escape). More antitops remain in the bags, with  more likely to be ``recycled" by the sphalerons: this produces more baryons than anti-baryons. Apparently we got the right sign for the baryon asymmetry.

Finally, let us comment on lepton CP violation. We do not know yet the corresponding CP-violating phase and the mixing angles in this sector,
so can lepton amplitudes similar to what we did with quarks produce larger effect? The factor $J$ above may be  larger: but the 
masses of the lower-component quarks $m_s,m_b$ would be changed to very small neutrino masses, with extremely small phases $\delta_i$
and much smaller overall CP violation.

\vskip .25cm {\bf Acknowledgments.} \vskip .2cm Our work  is
supported in parts by the US-DOE grant DE-FG-88ER40388.  Discussion with M.Shaposhnikov and J.Smit
of some related issues have been helpful to us at the onset of this work.

\end{document}